\documentclass[aps,prb,onecolumn,superscriptaddress,preprint]{revtex4-1}

\usepackage[]{graphicx} 
\usepackage{epsfig}
\usepackage{epstopdf}
\usepackage{float}
\usepackage{subfigure}
\usepackage{amsmath}
\usepackage[section]{placeins} 	
\usepackage{mathtools}

\usepackage{needspace}
\usepackage{color,soul}

\begin{document}


\title{Vibrational and thermoelastic properties of bcc iron from selected EAM potentials}


\author{Daniele Dragoni}
\altaffiliation[Present address:\ ]{Dipartimento di Scienza dei Materiali, Universit\`a di Milano-Bicocca, 20125, Milano, Italy}
\affiliation{Theory and Simulation of Materials (THEOS) and National Centre for Computational Design and Discovery of Novel Materials (MARVEL), \'Ecole Polytechnique F\'ed\'erale de Lausanne, CH-1015 Lausanne, Switzerland}

\author{Davide Ceresoli}
\affiliation{CNR Istituto di Scienze e Tecnologie Molecolari (CNR-ISTM) and INSTM UdR di Milano, 20133, Milano, Italy}

\author{Nicola Marzari}
\affiliation{Theory and Simulation of Materials (THEOS) and National Centre for Computational Design and Discovery of Novel Materials (MARVEL), \'Ecole Polytechnique F\'ed\'erale de Lausanne, CH-1015 Lausanne, Switzerland}


\date{\today}

\begin{abstract}
A comprehensive, critical study of the vibrational, thermodynamic and thermoelastic properties of bcc iron is presented, using well established semi-empirical embedded-atom method potentials available in the literature. 
Classical molecular dynamics simulations are used to address temperature effects, where dynamical matrices are constructed as a time average of the second 
moment of the atomic displacements. The $C_{11}, C_{44}, C'$ elastic constants are then obtained from the sound velocities along high symmetry directions in reciprocal space. 
Results are compared to ultrasonic measurements and highlight the limitations of the potentials considered here in describing thermoelastic properties.

\end{abstract}

\maketitle

\section{Introduction}

Elemental iron is a transition metal whose relative abundance in the universe is a by-product of stellar activity. It can be found in large quantities in liquid and solid forms in planetary cores like in the case of earth (85\% of the composition~\cite{earth_Fe}), thus characterizing the propagation of seismic waves in the interior of our planet.  It is also the major constituent of steels which are still of fundamental interest for our economic processes. 
The complexity of its phase diagram, mainly driven by the cooperative vibrational and magnetic contributions to the free energy~\cite{Pettifor,Koermann1}, makes this element particularly challenging to describe via computer simulations. 
This is valid especially at high temperatures, where these cooperative effects determine a dramatic change in the structural properties of the system. At the Curie temperature of  $\sim$ 1043~K, iron turns from a ferromagnet into a paramagnet experiencing a second order transition ($\alpha \to \beta$). 
This transition is then followed at higher temperatures by two other structural transitions, namely a bcc$\to$fcc ($\beta \to \gamma$) and a fcc$\to$bcc ($\gamma \to \delta$), before melting at $\sim$~1810~K.

Empirical potentials have been extensively tested and used to study the thermodynamic and mechanical properties of iron and its alloys, 
including phase stability and structural martensitic transitions~\cite{engin,muller,Lopasso-fe-cu,Bonny-fe-cr,Razumov2017}, and a vast class of point- or extended defects such as 
vacancies~\cite{ChuChunFu_FeH,malerba2010comparison}, interstitials~\cite{Olsson_Fe,Tapasa_FeC,malerba2010ab}, dislocations~\cite{HFe_carter,FeH_Curtin,Chaussidon_Fe,Osetsky_Fe,Yip,Li_FeH}, or 
tip-cracks with brittle to ductile transitions~\cite{DeVita-Crack,Kermode-brittle-fracture,brittletoductile}. The study of these classes of defects requires the use of accurate potentials that 
are capable of reproducing plastic, non-elastic and elastic properties at the same time. In fact, it is known that extended defects induce long-range residual stresses that can 
directly influence structure and dynamics of their own core-defect region~\cite{DeVita-Crack,Rao}. 
While zero-temperature experimental equilibrium volume and elastic constants are typically well reproduced by most of the EAM potentials available in the literature (due to the fact that, commonly, 
they are explicitly included in the fitting datasets together with other standard quantities such as lattice parameters, cohesive energies, and defect formation energies from experiments and calculations), 
there are only a few studies which have been performed to analyze carefully the performance and the accuracy of this kind of potentials at increasingly high temperatures~\cite{Adamsthermo,bian2008vibrational}. 
For iron in particular, we are not aware of such studies. 
For this reason, in this work we perform an extended investigation of the thermo-mechanical response of a selection of popular embedded-atom method (EAM) potentials~\cite{mendelev,meyer,ouyang,marchese} 
fitted on experimental and ab-intio data. The analysis is performed by means of classical molecular dynamics (MD) calculations throughout the entire range of stability of the $\alpha$-phase of iron, and 
can be considered a stringent test for the validation of a potential.
Very recently~\cite{Partay2018}, the high pressure/high temperature phase diagram of iron has been derived from a set of
EAM potentials, including some of the EAM potentials investigated in our study. The conclusion of Ref.~\onlinecite{Partay2018}
is that the accuracy of the EAM potentials depends strongly on the fitting set (i.e. including or not liquid configurations).

The paper is organized as follows: in Sec.~\ref{sec:methods} we describe the methodology used to calculate thermodynamic quantities, phonons and elastic constants. 
In Sec.~\ref{sec:checks} we analyze convergence issues associated to the methodology adopted for the calculation of the elastic constants.
The results are then compared with experimental data available in literature and discussed in Sec.~\ref{sec:results}. Summary and conclusions are reported in Sec.~\ref{sec:conclusions}.

\section{Methods and details of the calculations}
\label{sec:methods}
EAM potentials~\cite{EAM1} are a class of semi-empirical interatomic potentials constructed to provide an improved description of metallic bonding between atoms, compared to that of simple pair-wise 
interactions. 
The analytic functional form of such potentials shares similarities with that of the glue model~\cite{Gluemodel}, the Finnis-Sinclair~\cite{finnissinclair} approach and effective-medium 
theory~\cite{effmedtheory}. Specifically, the potential energy of an EAM atom $i$ embedded in a generic atomic environment is given by
\begin{align}
 \epsilon_i &= \frac{1}{2} \sum_{i\neq j}\phi(r_{ij}) + \mathcal{F} \Big( \sum_{i\neq j}\rho(r_{ij}) \Big),
 \label{EAM_eq}
\end{align}
where $r_{ij}$ is the distance between neighboring atoms $i$ and $j$ ($j$ within a sphere centered around $i$ with a local cutoff radius $r_{cut}$), $\phi(r)$ is a pair-wise potential term, 
and $\mathcal{F}$ a many-body convex nonlinear embedding function depending on an effective local charge density function $n(r_i)=\sum_{i\neq j}\rho(r_{ij})$ due to the atoms surrounding $i$ within 
a distance $r_{cut}$. 
Extensive reviews of the EAM approach can be found in Refs.~\onlinecite{EAMreview1,EAMreview2}. In this work, we focus our attention on the Mendelev03~\cite{mendelev}, Meyer98~\cite{meyer}, Ouyang12~\cite{ouyang} 
and Marchese87~\cite{marchese} EAM potential parameterizations, which have proven to be successful in the description of a wide range of crystal and defect properties of iron and its alloys
(cohesive energy, 0~K elasticity, mono-vacancy formation energy).
These potentials differ in the details of the functional form used to describe the embedding, in the pair and effective charge-density functions, and also in the different information included in the datasets.

In particular the Mendelev03 potential (\emph{potential 2} in
Ref.~\onlinecite{mendelev}) has been fit to both experimental data (including lattice spacing of bcc and fcc at 0~K, bcc cohesive energy, unrelaxed
bcc vacancy formation energy, bcc to fcc crystals energy difference, bcc and fcc
interstitial formation energies, liquid density) and to the forces obtained from a few snapshots
of an ab-initio MD simulation of liquid, non-magnetic iron.
The Meyer98 potential has been fit to various experimental data
sources, including bcc lattice constant, sublimation energy, elastic constants,
bcc-to-fcc energy differences, vacancy formation energy and selected phonon
frequencies.
The Marchese87 potential has been generated starting from an early work by
Finnis and Sinclair, and has been fit to experimental cohesive energy, equilibrium volume and elastic constants, in
addition to the potential energy along the vacancy migration barrier, but
they don't give details of the calculation.
Finally, the Ouyang12 EAM potential has been fitted to experimental data, including
bcc/bcc lattice parameters, elastic constants, cohesive energies, vacancy formation
energies in bcc/fcc, bond-length and dissociation energy of the Fe dimer.

Regardless of the parameterization, low-temperature experimental elastic constants of bcc iron have always been included in the training protocol. These potentials are therefore expected to reproduce such quantities 
at zero temperature and/or in the low-temperature regime. However, no systematic analysis has been performed to verify accuracy at finite temperature 
in the whole experimental temperature range of stability of the $\alpha$ (0$\rightarrow$1043~K) and $\beta$ (1043$\rightarrow$1185~K) phases, 1185~K being the experimental melting temperature.
In order to get a glimpse on the high-temperature behavior of the four potentials, we calculated their
melting temperature using the two-phase coexistence method.~\footnote{We prepared a 8000 atoms sample, with
two solid (bcc)-liquid interfaces, equilibrated at different temperature, and monitored the movement of
the solid-liquid interfaces. This way, we obtained a rough estimate of the melting temperature. Finally,
we performed a constant-enthalpy run and obtained the melting point as the average temperature of the system.}
The melting temperatures are the following: $1768\pm16$~K (Mendelev03), $2120\pm17$~K (Meyer03), $2404\pm20$~K (Marchese87) and $2276\pm22$~K (Ouyang12). From these results, the Mendelev03 potential, which has been
fitted also to the properties of the liquid, provides the best description of melting of solid iron.

The strategy adopted to calculate finite-temperature elastic constants is based on molecular dynamics (MD) simulations:
\begin{enumerate}
 \item We first compute the thermal expansion from constant pressure MD simulations, and extract the volumetric/linear thermal expansion coefficient $\alpha_V(T)/\alpha_L(T)$ and specific heat at constant 
 pressure $C_P(T)$.
\item Second, we calculate the phonon spectrum for a number of temperatures at their respective calculated equilibrium volumes, using the time average of the second moment of atomic displacements.
 \item Then, we compute the $C_{11}, C_{44}$ and $C'$ elastic constants as a function of temperature from the long-wavelength limit of the finite-temperature phonon dispersions. We derive the $C_{12}$ elastic 
 constant and bulk modulus $B$ from standard relationships for cubic crystals.
\end{enumerate}

We now describe these steps in detail. The equilibrium volumes are obtained performing a set of constant pressure/temperature (NPT) runs at vanishing external pressure, at a temperature going from 100 to 
1200~K with increments of 100~K. The pressure is controlled through a Parrinello-Rahman 
barostat~\cite{Parrinello-Rahman} while a Nose-Hoover chain thermostat~\cite{Nose-Hoover} is used to keep constant the average temperature. The equations of motion used to sample trajectories in the 
position-velocity phase-space of the NPT ensemble are those of Shinoda~\cite{Shinoda}, that combine the Martyna, Tuckerman and Klein correction~\cite{Martyna} with the strain energy proposed by Parrinello and
Rahman~\cite{Parrinello-Rahman}, and are solved using the time-reversible measure-preserving Verlet algorithm derived by Tuckerman~\cite{Tuckerman} as implemented in the LAMMPS~\cite{lammps} package.
The initial configuration of each MD run consists of a 10$\times$10$\times$10 cubic supercell with periodic boundary conditions (PBCs) containing 2000 atoms with slightly randomized displacements from the 
perfect bcc structure. 
The velocities are initialized according to a Maxwell-Boltzmann distribution. During the simulations, the time-step is fixed at 1~fs, and the relaxation times of the barostat and thermostats are set to 
be 1~ps and 0.1~ps, respectively. Each simulation is carried out for 10 million steps, equivalent to 10~ns. The first 0.5~ns are used for thermalization and equilibration of the system, while the remaining 
9.5~ns are used for accumulating thermodynamics averages. The simulation length and size are chosen to ensure the convergence of the relevant thermodynamic quantities. 
The volumetric and linear coefficients of thermal expansion are obtained from the temperature derivative of a cubic spline interpolation of the average equilibrium volumes calculated from the MD runs 
according to Eq.~\ref{eq:expcoeff_eq},
\begin{align}
 \alpha_V(T) &= 3\alpha_L(T)=\frac{1}{V} \frac{\partial V(T,P)}{\partial T} \Big|_{P=0} .
 \label{eq:expcoeff_eq} 
\end{align}
Similarly, according to the Eq.~\ref{eq:Cp_eq}, the specific heat is obtained as a temperature derivative of a cubic spline interpolation of the calculated average enthalpy $H$,
\begin{align}
 C_P(T)    &=\frac{\partial H(T,P)}{\partial T} \Big|_{P=0} .
 \label{eq:Cp_eq}
\end{align}

The phonon dispersions at finite temperature are obtained directly through MD runs using the \textit{FixPhonon} fix by Kong~\cite{kong} implemented 
into LAMMPS~\cite{lammps}. In this method, the dynamical matrix is obtained through Green's functions~\cite{kong2} calculated as time-averaged second moments of the atomic displacements, assuming  
thermal equilibrium and equipartition.
The eigenvalues of the dynamical matrix are computed at $\boldsymbol{q}~=~(\frac{ k_1}{N_1} \boldsymbol{b}_1, \frac{k_2}{N_2} \boldsymbol{b}_2, \frac{k_3}{N_3} \boldsymbol{b}_3)$ commensurate with 
the supercell size, where $N_i$ is the multiplicity of the primitive reciprocal space vector $\boldsymbol{b}_i$~\cite{Srivastava}, $k_i=0,\dots,N_i-1$. 
The acoustic sum rule is used to enforce the condition of zero frequency in $\Gamma$ as required by crystal translational symmetry. 
This approach accounts for anharmonic contributions that are naturally present in MD; these are recast into a standard dynamical matrix whose eigenvalues depend explicitly on the temperature. 
On the other hand, a strong harmonic behavior and long phonon lifetimes make this approach less efficient in sampling the atomic vibrational modes at low temperature.
It is also worth to point out that standard MD samples a classical Maxwell-Boltzmann probability distribution that deviates from the actual quantum Bose-Einstein one. 
This approach therefore neglects quantum effects like zero-point motion and the freezing out of the high-energy vibrations at low temperature. 
Consequently, in practice, even though the whole range of experimental thermodynamic stability of $\alpha$-$\beta$ iron is  accessible via MD simulations, caution is recommended in comparing calculations
and experiments well below the Debye temperature $\Theta_D$ (from experiments~\cite{Kittel}  $\Theta_D~\sim$~480~K). 

The phonons are computed using a constant volume-temperature (NVT) ensemble from 200~K to 1200~K with increments of 200~K. The supercell volumes are adjusted to match those obtained by the 
thermal expansion curves calculated previously. The temperature is controlled by the Nose-Hoover chain technique, described in the previous paragraphs. In order to access directly the phonon dispersion in 
the bcc BZ, we use a $N\times N\times N$ bcc supercell rather than a cubic one.   
The choice of the supercell size, i.e. the $N$ value, as well as the simulation length are crucial for a good convergence of the phonon frequencies, especially around $\Gamma$, and are exhaustively 
and separately discussed in Sec.~\ref{sec:checks}.

From the phonon dispersions, we obtain the sound velocities by calculating the slope of the dispersion branches along high-symmetry directions in the first BZ (see Sec.~\ref{sec:checks} for details). 
Then, we derive the elastic constants by inverting of their relationships with the sound velocities of elastic waves as from Ref.~\onlinecite{Srivastava},
\begin{equation} 
 \begin{aligned}
 v_L \left[ 100 \right] &=\sqrt{\frac{C_{11}}{\rho}},          \\
 v_{T_1} \left[ 100 \right] &=\sqrt{\frac{C_{44}}{\rho}},      \\
 v_{T_2} \left[ 110 \right] &=\sqrt{\frac{C'}{\rho}} ,          
 \label{eq:cost}
 \end{aligned}
\end{equation}
where $\rho$ is the mass density, $v_L[100]$ is the acoustic longitudinal sound velocity propagating along the $[100]$ symmetry direction in a cubic crystal and $v_{T_1}[100]$, $v_{T_2}[110]$ are the acoustic 
transverse degenerate TA$_1$ and transverse non-degenerate TA$_2$ sound velocities along the [100] and [110] directions respectively.
The combined temperature dependence of the sound velocities and of the density $\rho$ provides the thermal behavior of the elastic constants. Moreover, since three elastic constants are sufficient to determine 
by symmetry the full stiffness tensor of a cubic crystal, we also derive the bulk modulus $B=(C_{11}+2C_{12})/3$ and the $C_{12}=C_{11}-2C'$ constant.

\section{Sound velocity calculations}
\label{sec:checks}
In this section, we describe the approach used to calculate elastic constants from phonon dispersions and analyze finite -time and finite-size effects.

We start performing NVT runs with a $8\times8\times8$ primitive bcc simulation box. Even though this is an arbitrary choice for the cell size (it will be discussed in more detail below), it provides 
a reasonable starting point to study the time-convergence of the vibrational spectrum. 
After thermalization and equilibration, we compute the dynamical matrix and the phonon dispersions every 1~ns by block-averaging the second moment atomic displacements. We then check the time convergence 
of the phase velocities  $\omega_{L}(\boldsymbol{q}_N^{[100]})/q_N^{[100]}$, $\omega_{T_1}(\boldsymbol{q}_N^{[100]})/q_N^{[100]}$, $\omega_{T_2}(\boldsymbol{q}_N^{[110]})/q_N^{[110]}$ 
along the high symmetry directions of interest. In these formulae $\omega_{L/T}$ are the temperature dependent phonon frequencies evaluated for the longitudinal/transverse branches at the reciprocal vectors
\begin{equation}
 \begin{aligned}
\boldsymbol{q}_N^{[100]} &= \frac{(-\boldsymbol{b}_1+\boldsymbol{b}_2+\boldsymbol{b}_3)}{N},  \\
\boldsymbol{q}_N^{[110]} &= \frac{\boldsymbol{b}_3 }{N} ,
 \end{aligned}
 \label{eq:q_N}
\end{equation}
associated to the maximum wavelength allowed for the supercell considered in the selected crystal direction; $q_N^{[100]}$, $q_N^{[110]}$ are the corresponding moduli.
From the analysis of the data, we observed that the autocorrelation time of the calculated phase velocities is quite long (especially at low temperatures) and a simulation time of 100~ns would
be necessary to reduce the standard deviation of the phase velocities to a value equivalent (from error propagation theory) to a few GPa in the elastic constant. 
In order to decrease the uncertainty of the statistical quantities under consideration, we run several (7) parallel independent runs (differing only in the initial random seed used for the velocities initialization).

Next, we consider supercell-size effects. We repeat the approach described above for the calculation of the phonon frequencies. We run for 100~ns, considering different supercell sizes with $N$ ranging 
from 4 to 20 in steps of 4, and a few test temperatures. 
Larger values of $N$ provide a denser sampling of the phonon spectrum in the first BZ. As $N\rightarrow \infty$, the $\boldsymbol{q}_N$ points defined in Eqs.~\ref{eq:q_N} get closer and closer 
to $\Gamma$ and their phase velocities also converge to the group velocities in $\Gamma$ i.e., to the speed of sound.

\begin{figure}[ht]
\centering{
  \begin{tabular}{c}
   \includegraphics[trim=0mm 0mm 0mm 0mm, clip, width=0.48\textwidth]{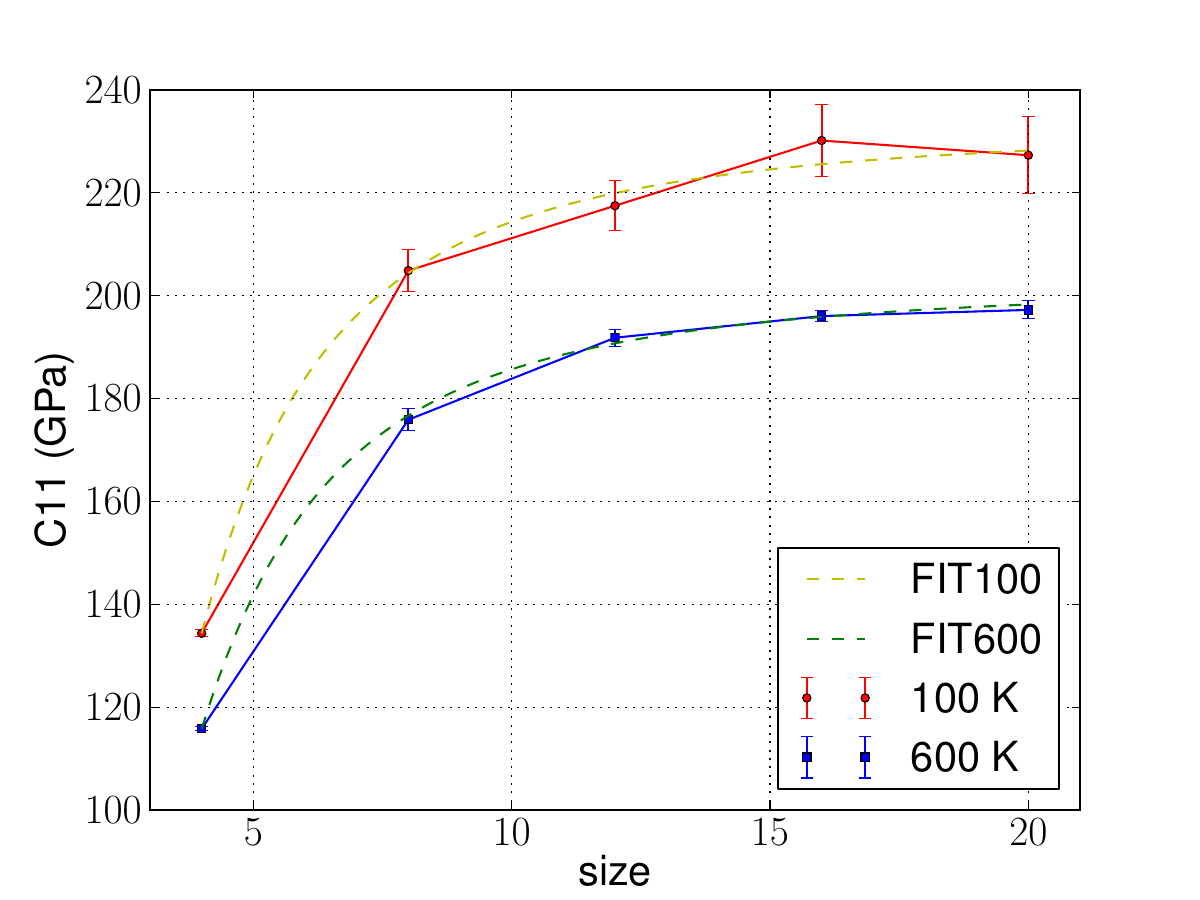} \\
   \includegraphics[trim=0mm 0mm 0mm 0mm, clip, width=0.48\textwidth]{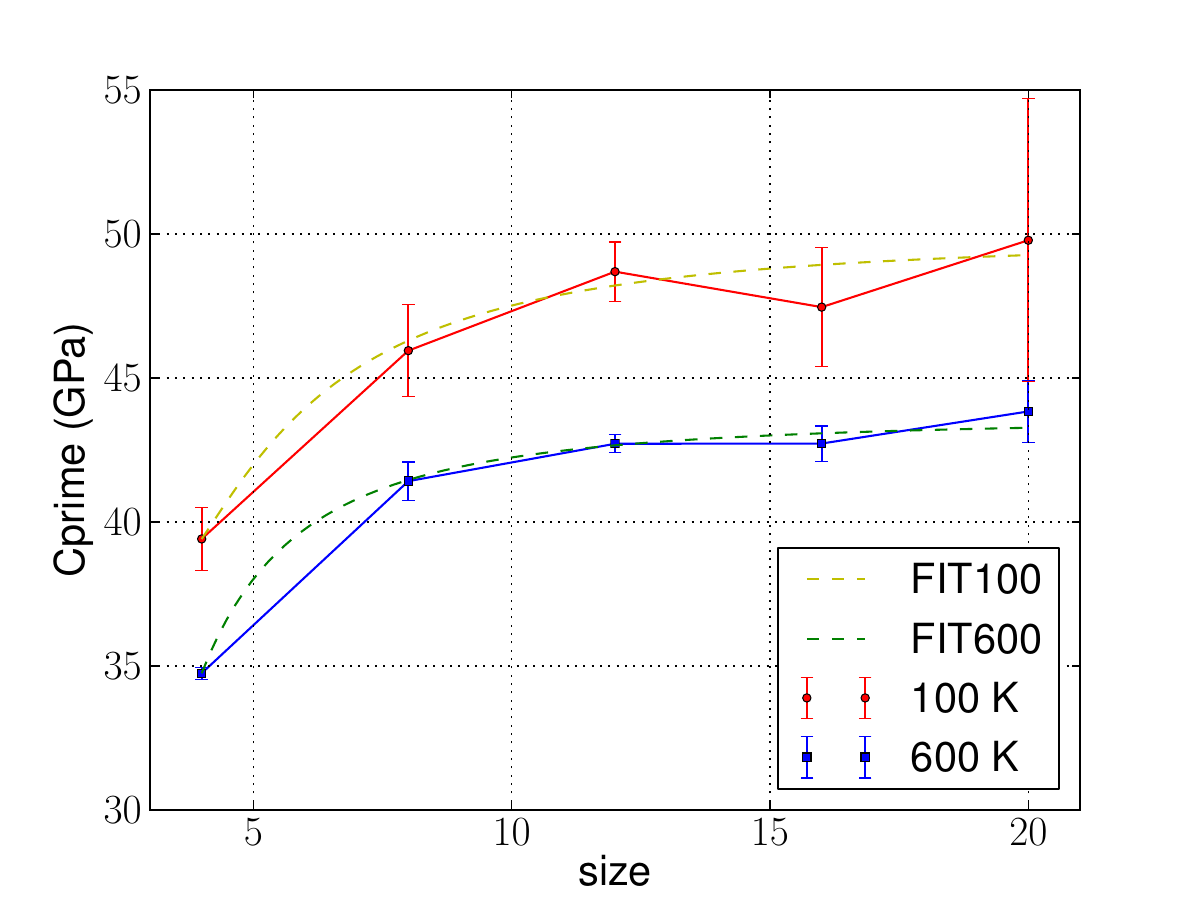} 
  \end{tabular}
 }
 \caption[...]{Convergence of the elastic constant $C_{11}$ (top panel) and of the elastic constant $C'$ (bottom panel) with respect to the size $N$ of the supercell for the 
 Mendelev03 potential at 100~K and 600~K respectively. Points with error bars, connected by straight lines, are the calculations at finite $N$. The dashed lines are the optimal fit for the temperatures 
 of interest performed with a model function for the sound velocity of the form $A_1+A_2/N^2+A_3/N^4$ as described in the text.
 }
 \label{fig:CON11}
\end{figure}

Since the acoustic phonon branches can be described as a superposition of sines of the wavevector, we expand $\omega(q_N)$ for small $q_N=|\boldsymbol{q}_N|$ (i.e. large $N$) as a 
one dimensional function along the symmetry direction of interest: 
\begin{equation}
\begin{aligned}
\omega(q_N) &\propto q_N - \frac{q_N^3}{3!} + \frac{q_N^5}{5!} + \mathcal{O}(q_N^7).\\
\end{aligned}\label{eq:fitt}
\end{equation}
It follows that the phase velocities in any direction can be rewritten as an explicit function of the supercell size $N$, taking the form $A_1+A_2/N^2+A_3/N^4$, where $A_1, A_2$ and $A_3$ are adjustable 
coefficients that depend on the specific phonon branch and direction in the first BZ and are to be determined. In fact, these are obtained fitting the phase velocities calculated at finite cell sizes. 
For any direction and phonon branch, the optimal fit is used to extrapolate the $N\rightarrow\infty$ limit of the phase velocity, i.e. the sound velocity (see for instance Fig.~\ref{fig:CON11}). 
By means of Eqs.~\ref{eq:cost}, this provides a measure of the finite size error made on the elastic constants for the different cell sizes.
This analysis suggests that elastic constants are well converged within the uncertainty due to the finite simulation time used in our MD runs for the case of a $20 \times 20 \times 20$ supercell. 
For this reason, we use this specific cell size and a simulation time of 100~ns (averaging over seven independent trajectories) as described 
above in order to calculate all the elastic constants for all the potentials at every temperature.

\section{Results and Discussion}
\label{sec:results}

\begin{figure}[ht]
\centering{
  \begin{tabular}{c}
   \includegraphics[trim=0mm 0mm 0mm 0mm, clip, width=0.46\textwidth]{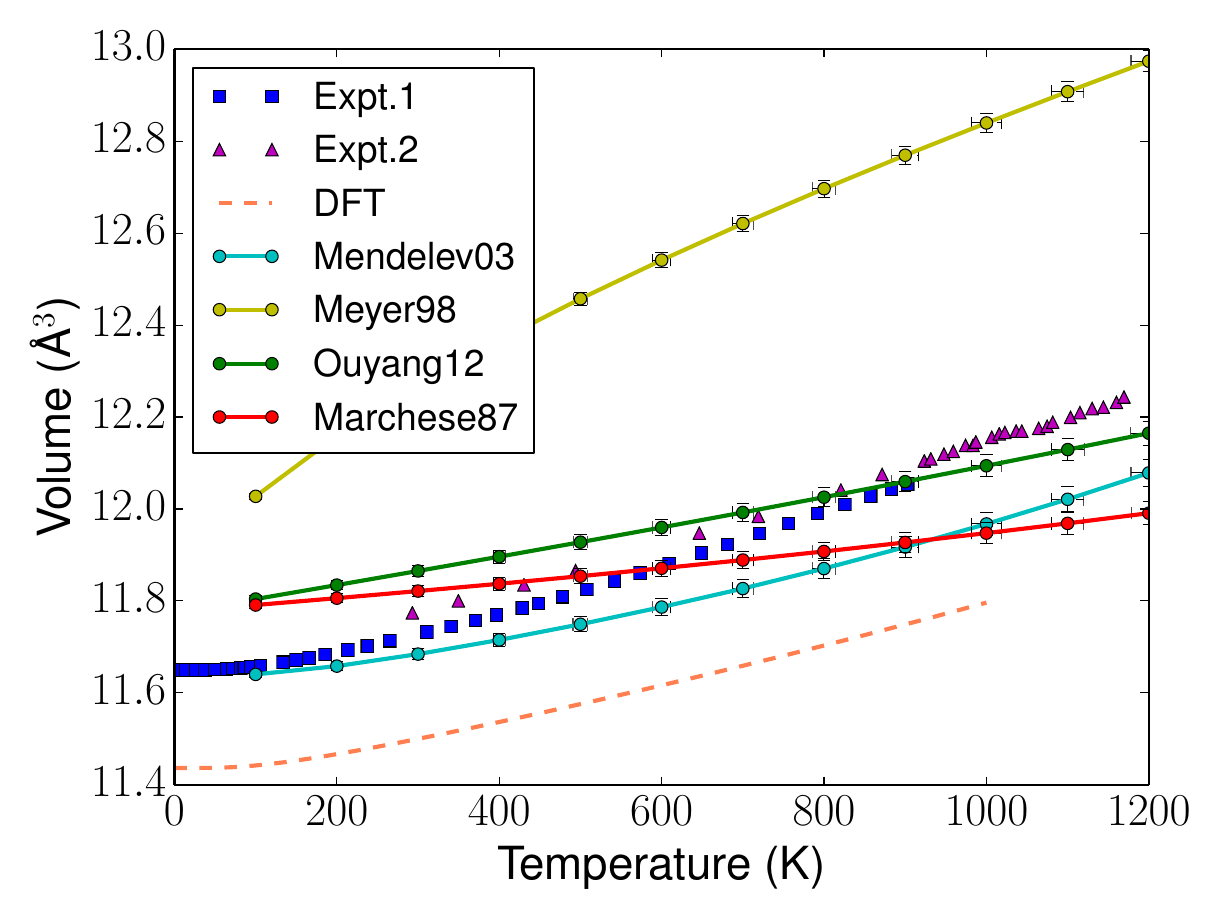} 
  \end{tabular}
 }
 \caption[...]{ Volumetric thermal expansion in the range of experimental thermodynamic stability of $\alpha$ and $\beta$ phases of iron for each of the empirical potential considered in this work. The results 
 are reported along with their standard deviations and compared to experimental data from Ref.~\onlinecite{Basinski} (Expt.1 -- blue squares), Ref.~\onlinecite{Ridley} (Expt.2 -- magenta triangles) and to 
 quasi-harmonic ab-initio data (dashed line) from Ref.~\onlinecite{Dragoni}.}
 \label{fig:therm_exp}
\end{figure}

We first compute the volume thermal expansion of the four EAM potentials considered in this work. The results are displayed in Fig.~\ref{fig:therm_exp} starting from 100~K up to 1200~K. Simple polynomial 
extrapolations of the EAM curves down to 0~K would provide values in good agreement with the different experimental data used to fit the various potentials (see Tab.~\ref{tab:0K-volumes}).
%
\begin{figure}[ht]
\centering{
  \begin{tabular}{c}
   \includegraphics[trim=0mm 0mm 0mm 0mm, clip, width=0.46\textwidth]{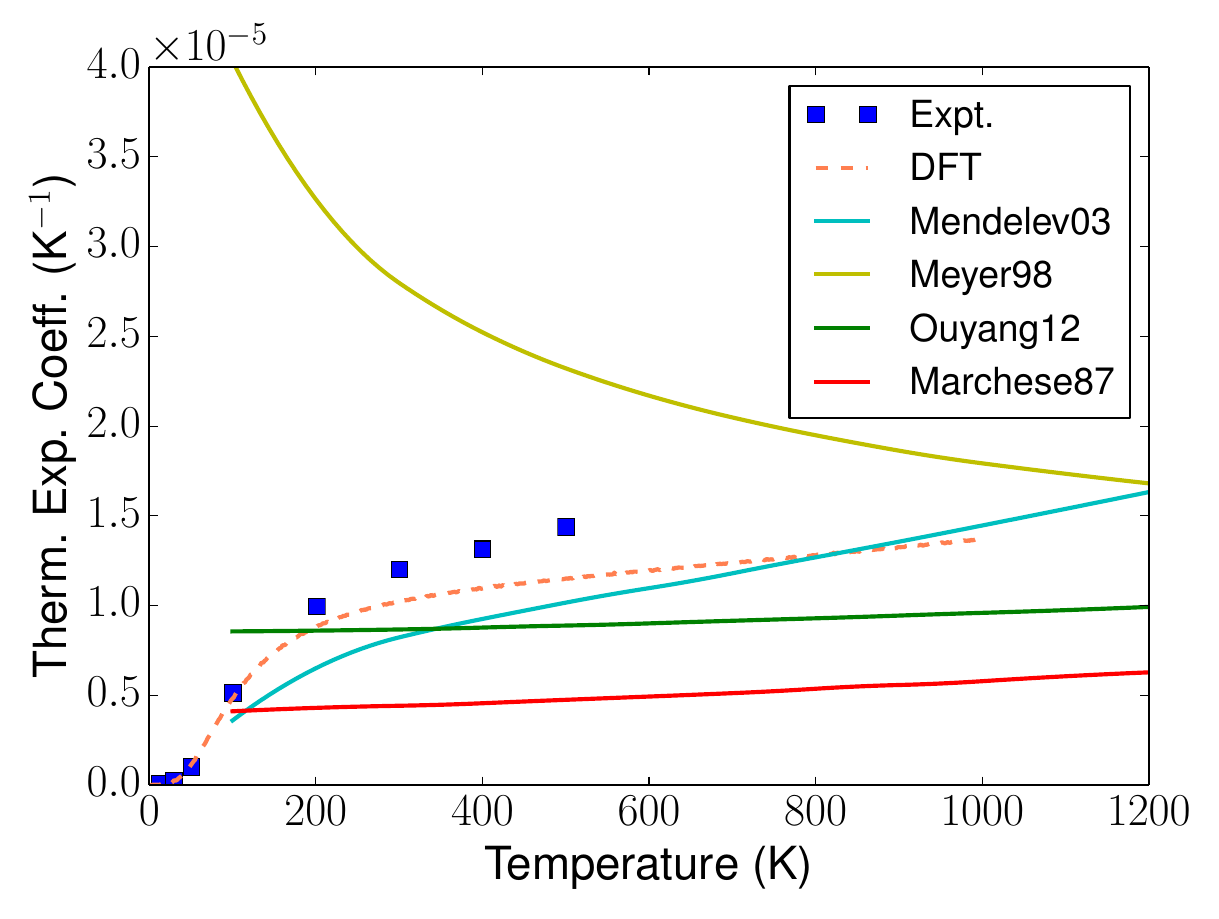} 
  \end{tabular}
 }
 \caption[...]{ Linear thermal expansion coefficient $\alpha_L(T)$ in the range of experimental thermodynamic stability of $\alpha$ and $\beta$ phases of iron for each of the empirical potential considered 
 in this work. As discussed in the text, these results are obtained as numerical derivatives of a cubic spline of the data in Fig.~\ref{fig:therm_exp} and are compared to experimental data from 
 Ref.~\onlinecite{Grigoriev} (Expt. -- blue squares) and to quasi-harmonic ab-initio data (dashed line) from Ref.~\onlinecite{Dragoni}.}
 \label{fig:therm_exp_coeff}
\end{figure}
\begin{table}
\centering
\begin{tabular}{c|@{\hspace{0.2cm}}cccc}
\hline\hline
              & Marchese87 & Mendelev03 & Meyer98 & Ouyang12 \\
\hline
V ($\AA^3$)   & 11.777 & 11.639 & 11.797 & 11.774 \\
$C_{11}$(GPa) & -- & 243.4 & 251.0 & 233.0  \\
$C'$  (GPa)   & -- & 49.3 & 60.3 & 48.8     \\
$C_{44}$(GPa) & -- & 116.0 & 118.7 & 117.8  \\
\hline\hline
\end{tabular}
\caption{0~K equilibrium volumes and elastic constants as reported in the original works of the different potentials. The volume value of the Marchese87 potential instead is calculated.}
\label{tab:0K-volumes}
\end{table}
\begin{figure}[ht]
\centering{
  \begin{tabular}{c}
   \includegraphics[trim=0mm 0mm 0mm 0mm, clip, width=0.46\textwidth]{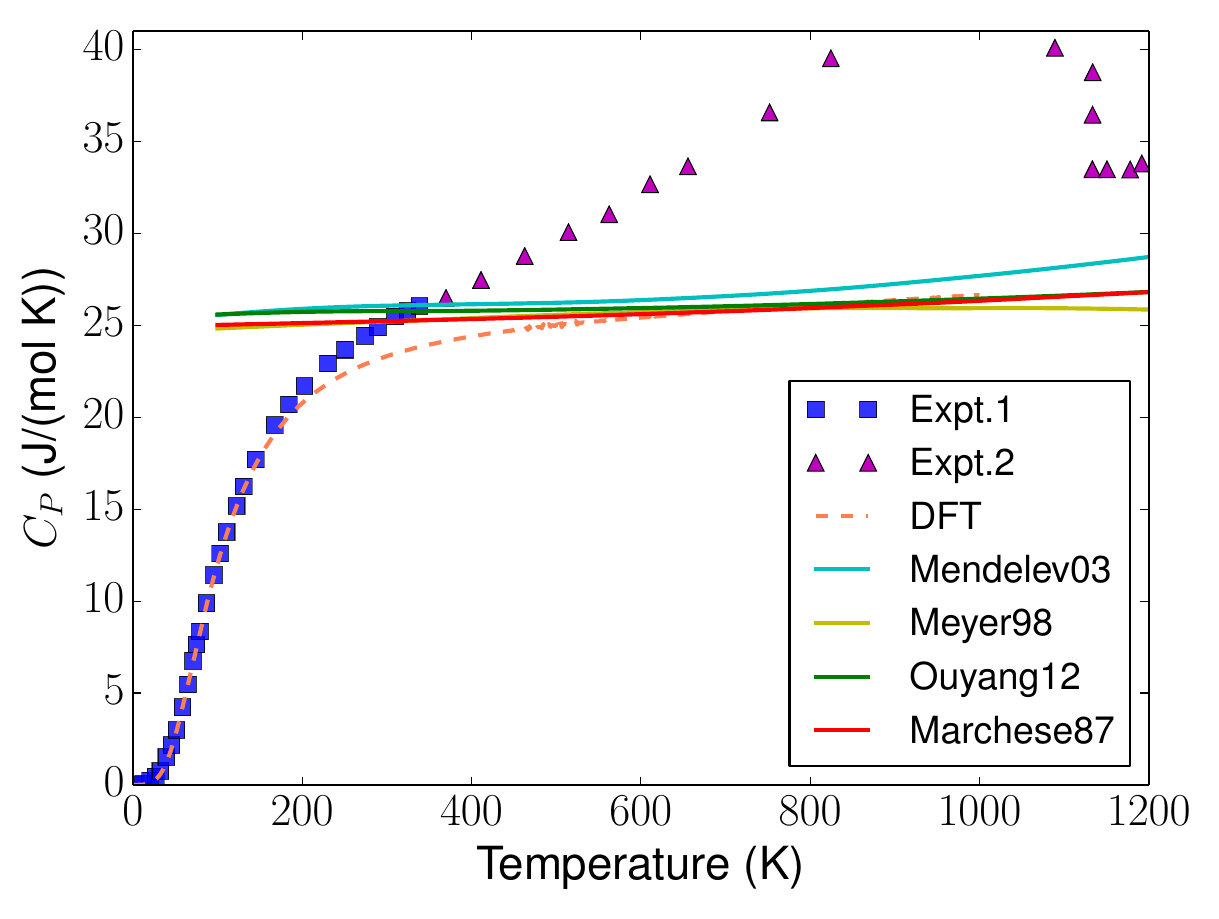} 
  \end{tabular}
 }
 \caption[...]{Heat capacity at constant pressure as a function of temperature for each of the empirical potential considered  in this work. 
 As discussed in the text, these results are obtained as numerical derivatives of a cubic spline of the Enthalpy data and are compared to experiments from 
 Ref.~\onlinecite{Desai} (Expt.1 -- blue squares), from Ref.~\onlinecite{Wallace2} (Expt.2 -- magenta triangles) and to quasi-harmonic ab-initio data (dashed line)
 from Ref.~\onlinecite{Dragoni}.}
 \label{fig:Cp_fig}
\end{figure}
\begin{figure*}[!ht]
\centering
\begin{tabular}{cc}
\centering  
\subfigure[]{ \includegraphics[width=8cm]{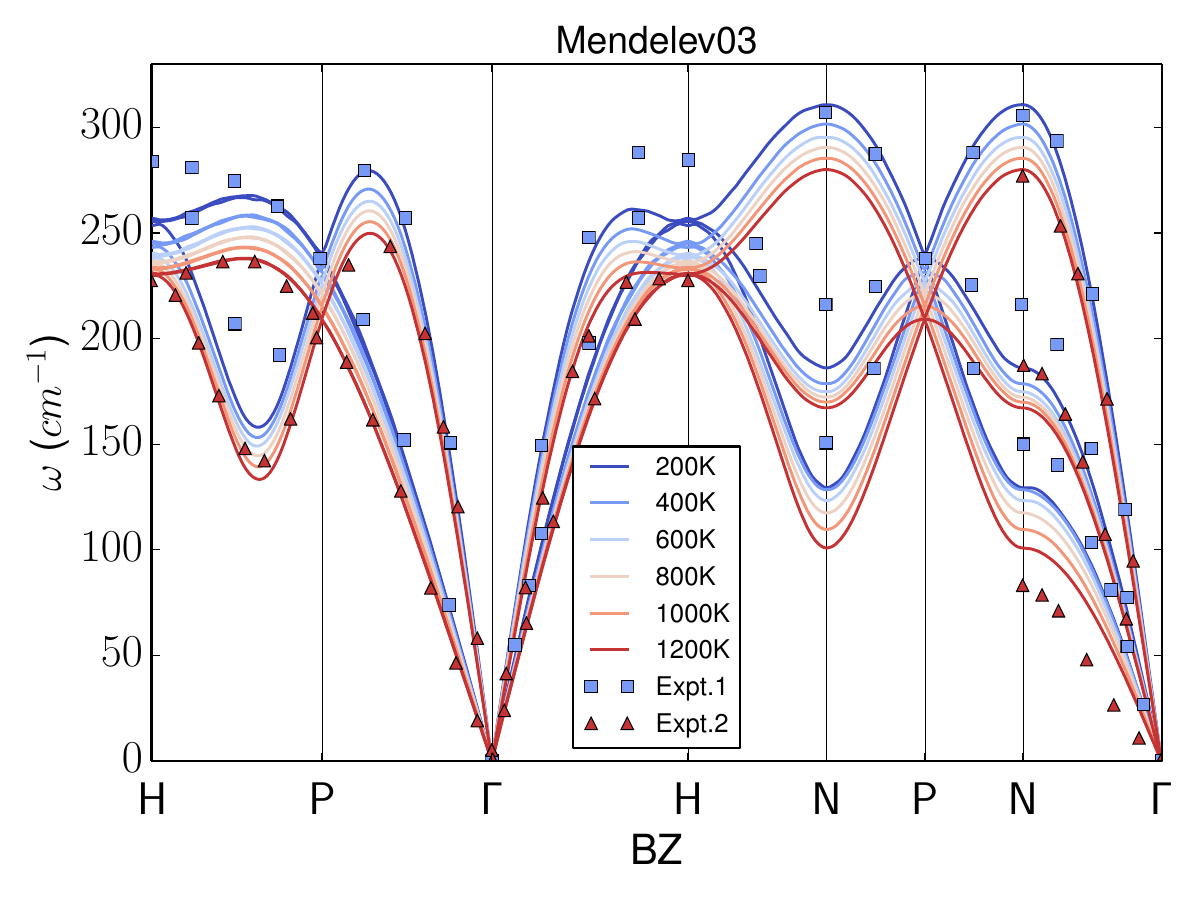}} &
\subfigure[]{ \includegraphics[width=8cm]{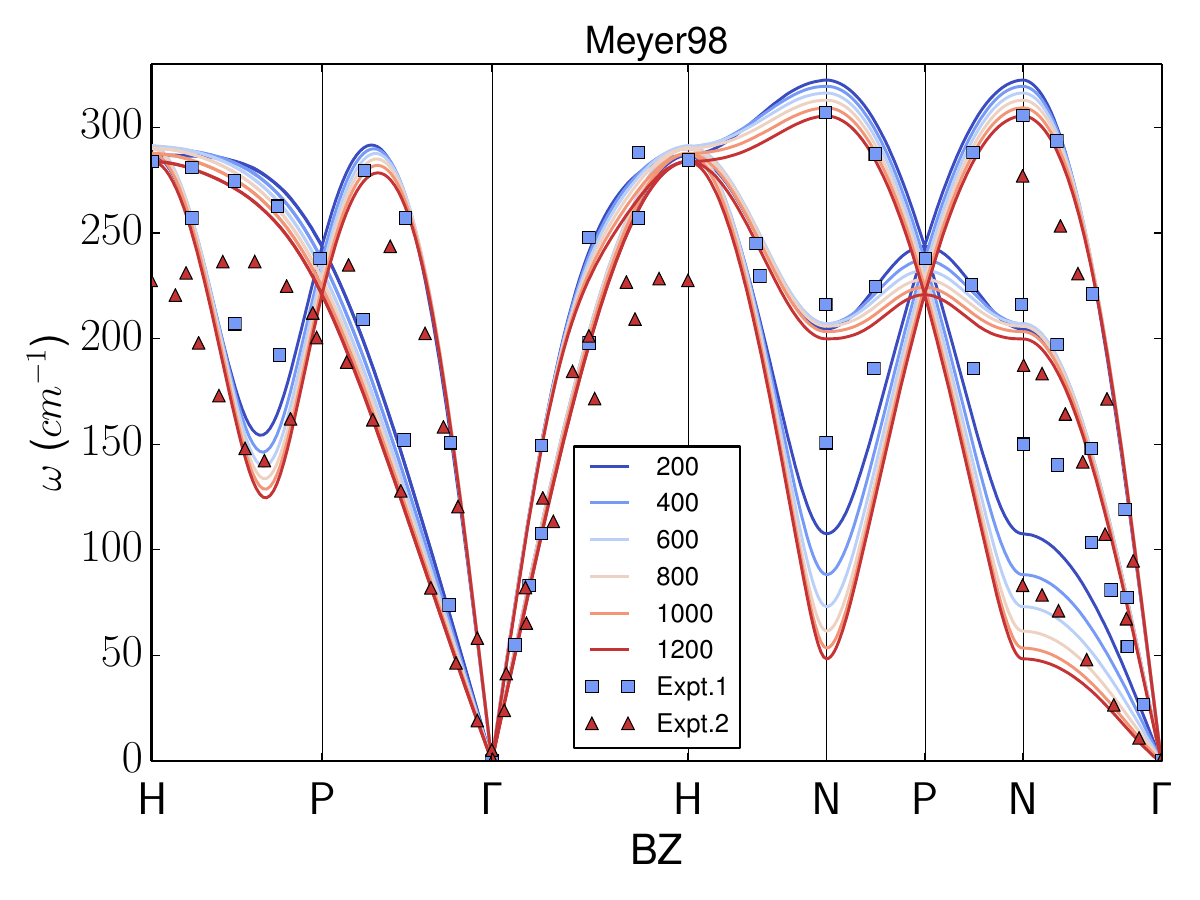}}    \\
\subfigure[]{ \includegraphics[width=8cm]{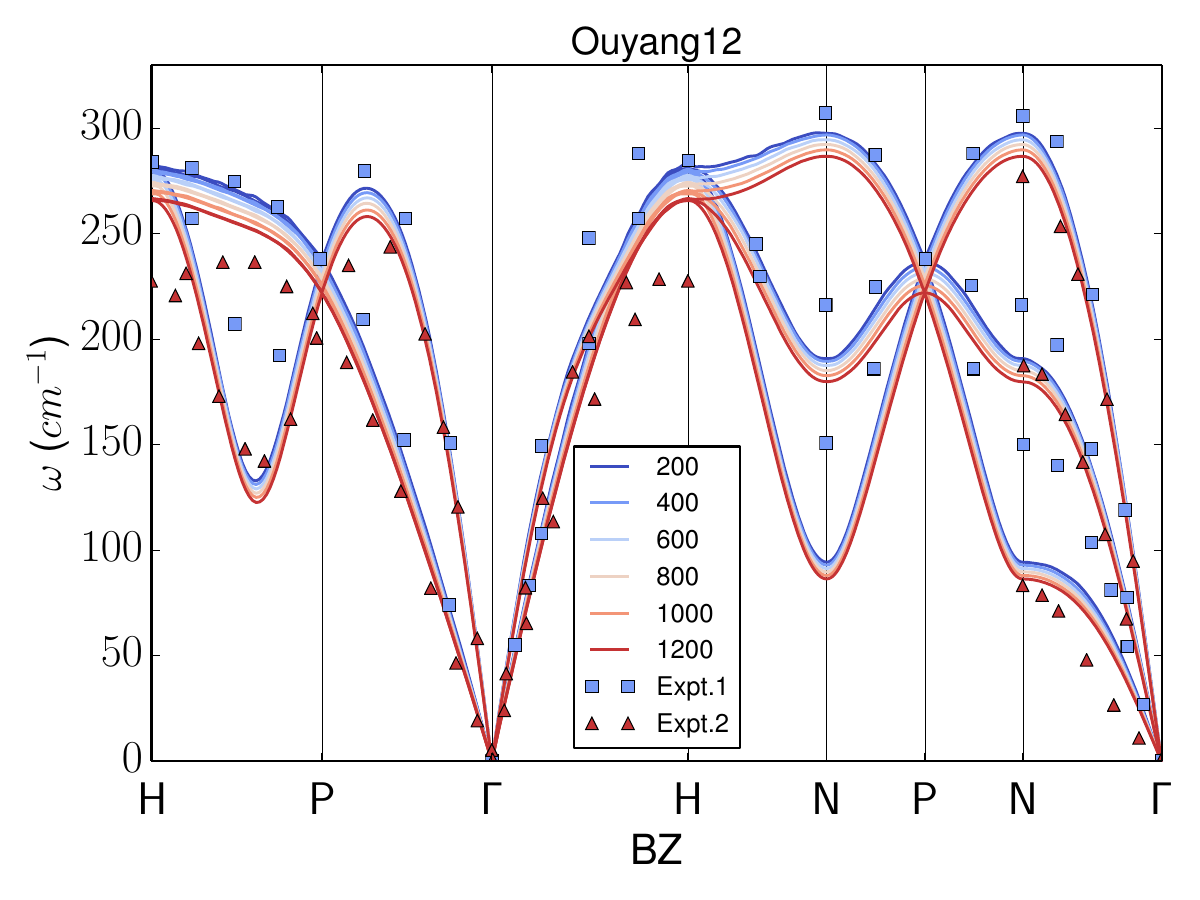}}   &
\subfigure[]{ \includegraphics[width=8cm]{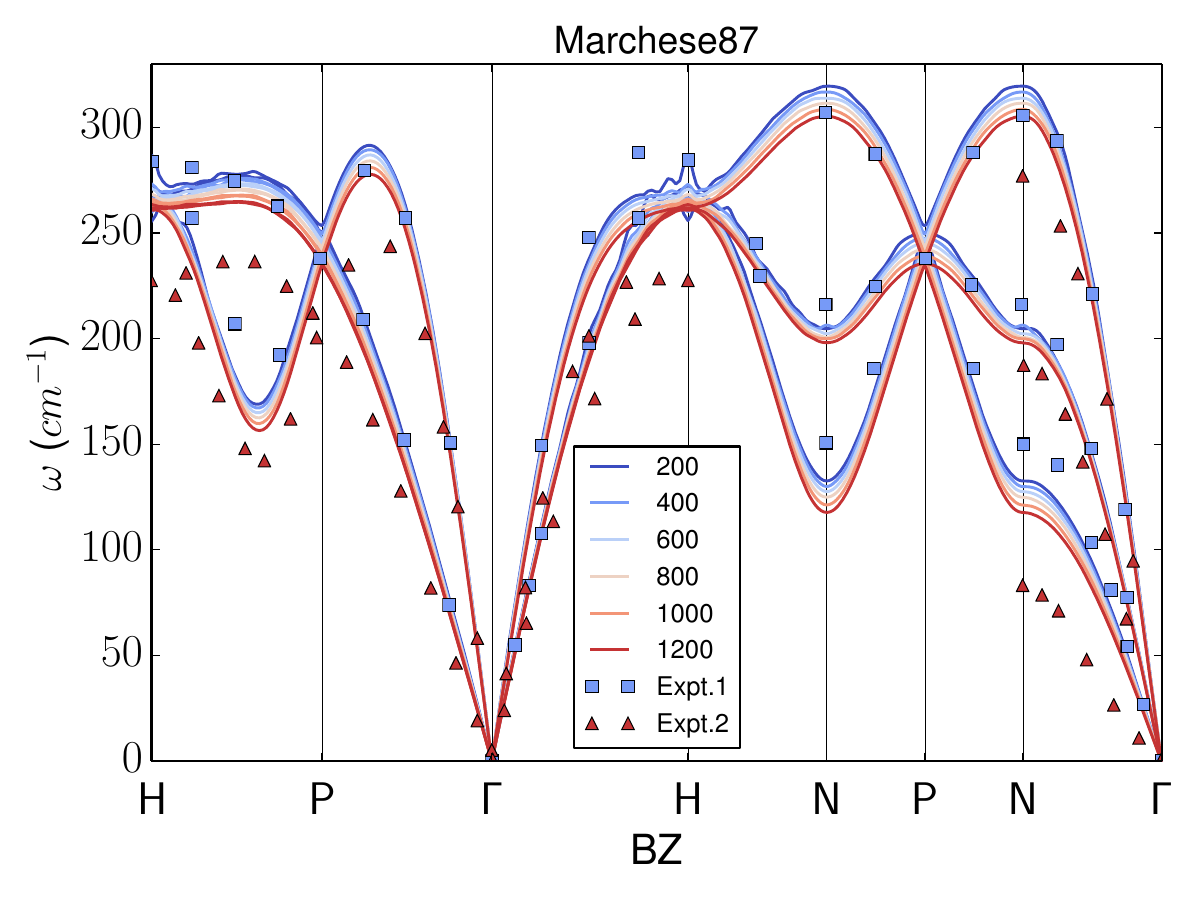}} \\
\end{tabular}
\caption{Temperature dependence of the phonon spectra of the four EAM potentials considered in this work (the potential name is specified at top of each graph). 
The continuous spectra are calculated as described in the main text at 6 different temperatures from 200 to 1200~K in steps of 200~K. The calculations are compared with experimental results from 
Ref.~\onlinecite{Brockhouse} measured at 300~K (Expt.1 -- squares) and Ref.~\onlinecite{Fultz} measured at 1158~K (Expt.2 -- triangles). 
}
\label{fig:PH_temperature}
\end{figure*}
%
Despite the expected agreement with the ground state equilibrium volume, the thermal expansion of most EAM potentials shows deviations from the experimental behavior. These deviations are more evident 
by looking at the thermal expansion coefficient curves reported in Fig.~\ref{fig:therm_exp_coeff}. Both the Marchese87 and Ouyang12 potentials show a linear, almost negligible, temperature dependence 
of the expansion coefficients across the temperature range of interest, displaying also absolute values that constantly underestimate the experimental data above 200~K.
The Meyer98 potential shows a peculiar monotonically decreasing coefficient that contrasts the experimental evidence both from a quantitative and a qualitative point of view.
The Mendelev03 potential, on the other hand, is the only one that displays a qualitatively correct behavior, replicating also the low temperature experimental trend. It is however worth to mention that
the low temperature experimental behavior is known to be dominated by quantum (Bose-Einstein) statistical effects that are not present in classical MD. 
The inclusion of such effects in the thermal expansion analysis of the classical potentials is therefore expected to introduce corrections to the classical picture presented in 
Fig.~\ref{fig:therm_exp} and Fig.~\ref{fig:therm_exp_coeff}. Additionally, the importance of such corrections is expected to increase as the zero-temperature limit is approached (a quantitative 
estimate for the zero point correction on the lattice from DFT quasi-harmonic analysis is provided in Ref.~\onlinecite{Dragoni}).

The heat capacity at constant pressure $C_P(T)$ is also calculated. The results from classical MD are reported in Fig.~\ref{fig:Cp_fig} and display an almost constant behavior with the absolute values close 
to 24.94\,J\,mol$^{-1}$K$^{-1}$($3\,R$), in accordance with the classical Dulong-Petit law (valid more specifically for the constant volume heat capacity $C_V$). 
As discussed above, the deviation from experimental data (and from the quasi-harmonic approximation data of Ref.~\onlinecite{Dragoni}) at low temperatures is a signature of classical statistics 
effects on vibrations. 
If we consider the high-temperature limit instead, the experimental data show a rapid increase with a divergence around 1000~K. As discussed in Refs.~\onlinecite{Neuge_C_P,Lavrentiev,Ruban} this is 
associated to the magnetic $\alpha$ to $\beta$ transition (Curie point 1043~K) and is due to the contribution of magnetic moments to the entropy of the system, and, to a lesser extent, to electronic 
excitations. Since neither magnetic nor electronic degrees of freedom are considered in classical MD, any deviation from the $3\,R$ limit in the high-temperature regime can than be ascribed to 
phonon-phonon anharmonic contributions.

%
%
\begin{figure}[!ht]
\centering{
  \begin{tabular}{c}
   \includegraphics[trim=0mm 0mm 0mm 0mm, clip, width=0.46\textwidth]{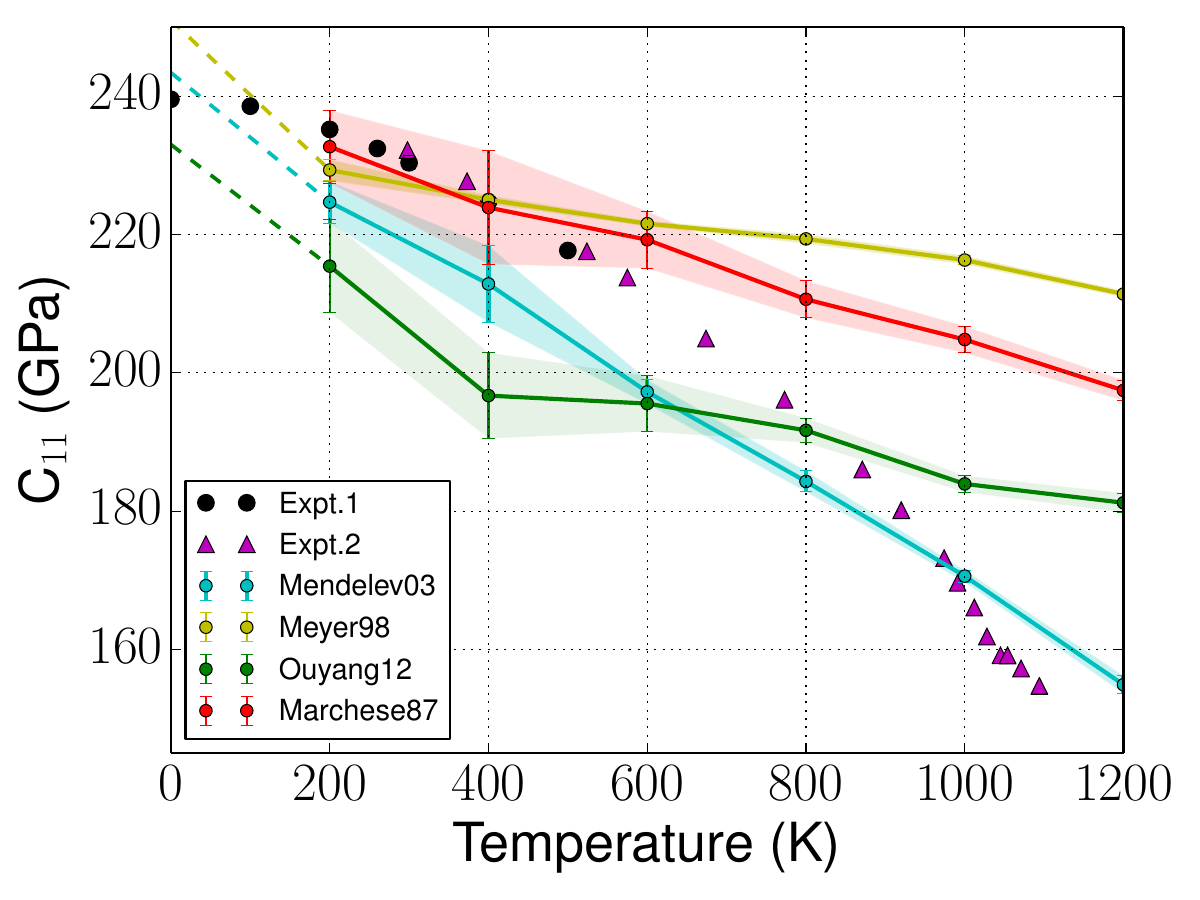} \\
  \end{tabular}
 }
 \caption[...]{  Temperature dependence of the  $C_{11}$ elastic constants obtained from the different EAM potentials. The elastic constants are calculated at discrete values of temperature and connected 
 by straight lines to provide continuous curves. Their confidence interval is also reported and highlighted by the shaded areas.  The dashed lines are guides to the eye that connect our curves with the 
 calculated 0~K values reported in the original works of the different potentials (where available) and collected in Tab.~\ref{tab:0K-volumes}.
 Our results are compared to ultrasonic experimental data from Ref.~\onlinecite{JJAdams} (Expt.1 -- black circles) and  Ref.~\onlinecite{Dever} (Expt.2 -- magenta triangles).
 }
 \label{fig:C11_TS}
\end{figure}

\begin{figure}[!ht]
\centering{
  \begin{tabular}{c}
   \includegraphics[trim=0mm 0mm 0mm 0mm, clip, width=0.46\textwidth]{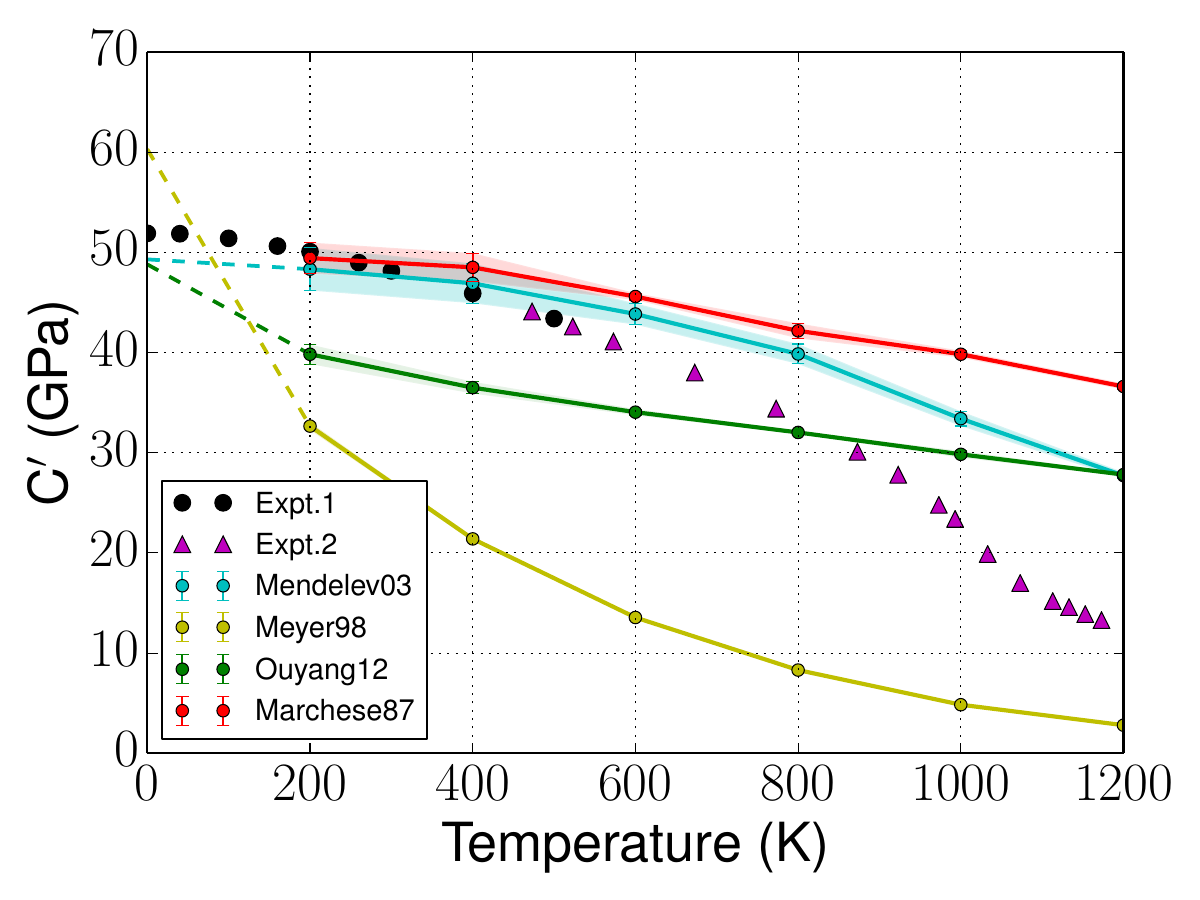} 
  \end{tabular}
 }
 \caption[...]{Temperature dependence of the $C'$ elastic constant from the different EAM potentials. The results are compared to ultrasonic experimental values from Ref.~\onlinecite{JJAdams} 
 (Expt.1 -- black circles) and Ref.~\onlinecite{Dever} (Expt.2 -- magenta triangles). Shaded areas and dashed lines as in Fig.~\ref{fig:C11_TS}.}
 \label{fig:Cp}
\end{figure}
\begin{figure}[!ht]
\centering{
  \begin{tabular}{c}
   \includegraphics[trim=0mm 0mm 0mm 0mm, clip, width=0.46\textwidth]{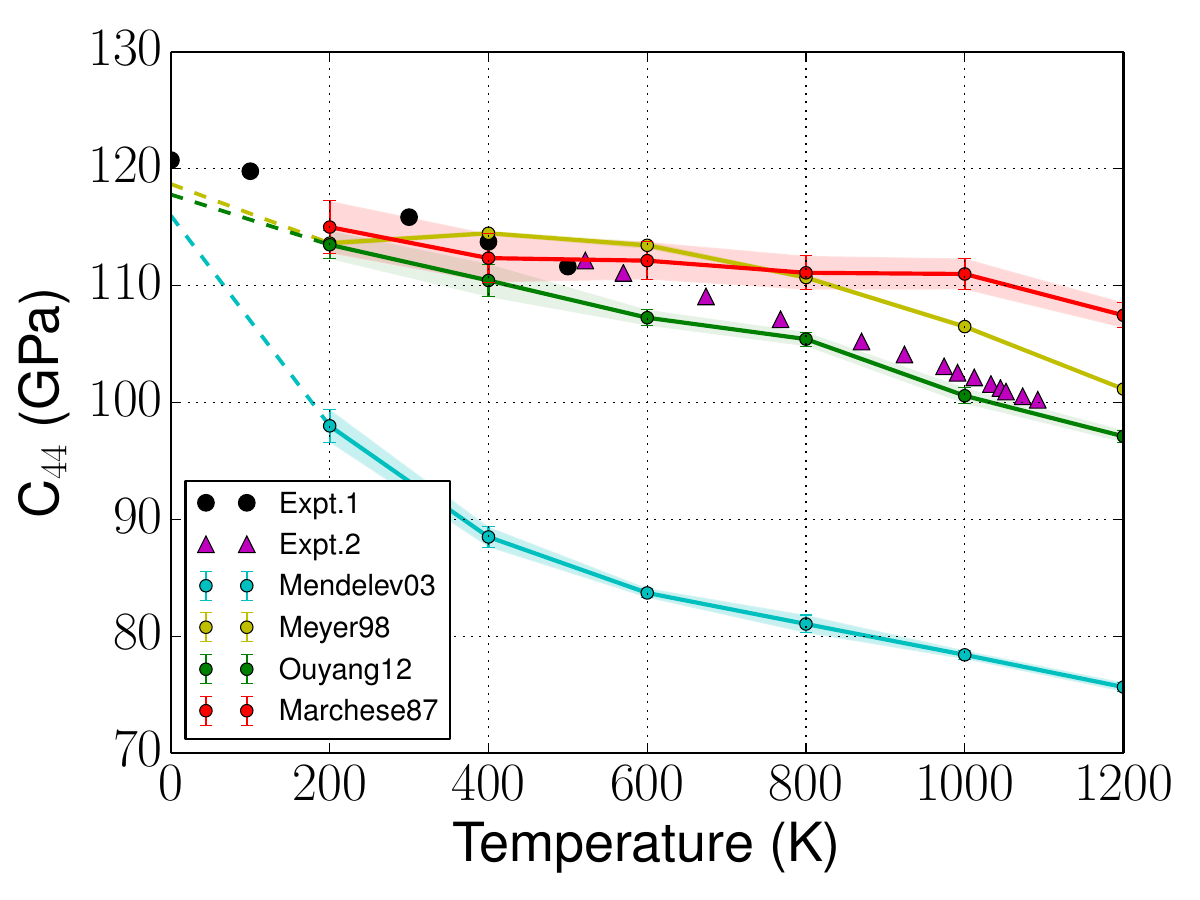} 
  \end{tabular}
 }
 \caption[...]{Temperature dependence of the $C_{44}$ elastic constant obtained from the different EAM potentials. The results are compared to ultrasonic experimental values from Ref.~\onlinecite{JJAdams} 
 (Expt.1 -- black circles) and Ref.~\onlinecite{Dever} (Expt.2 -- magenta triangles). Shaded areas and dashed lines as in Fig.~\ref{fig:C11_TS}.}
 \label{fig:C44}
\end{figure}
\begin{figure}[!ht]
\centering{
  \begin{tabular}{c}
   \includegraphics[trim=0mm 0mm 0mm 0mm, clip, width=0.46\textwidth]{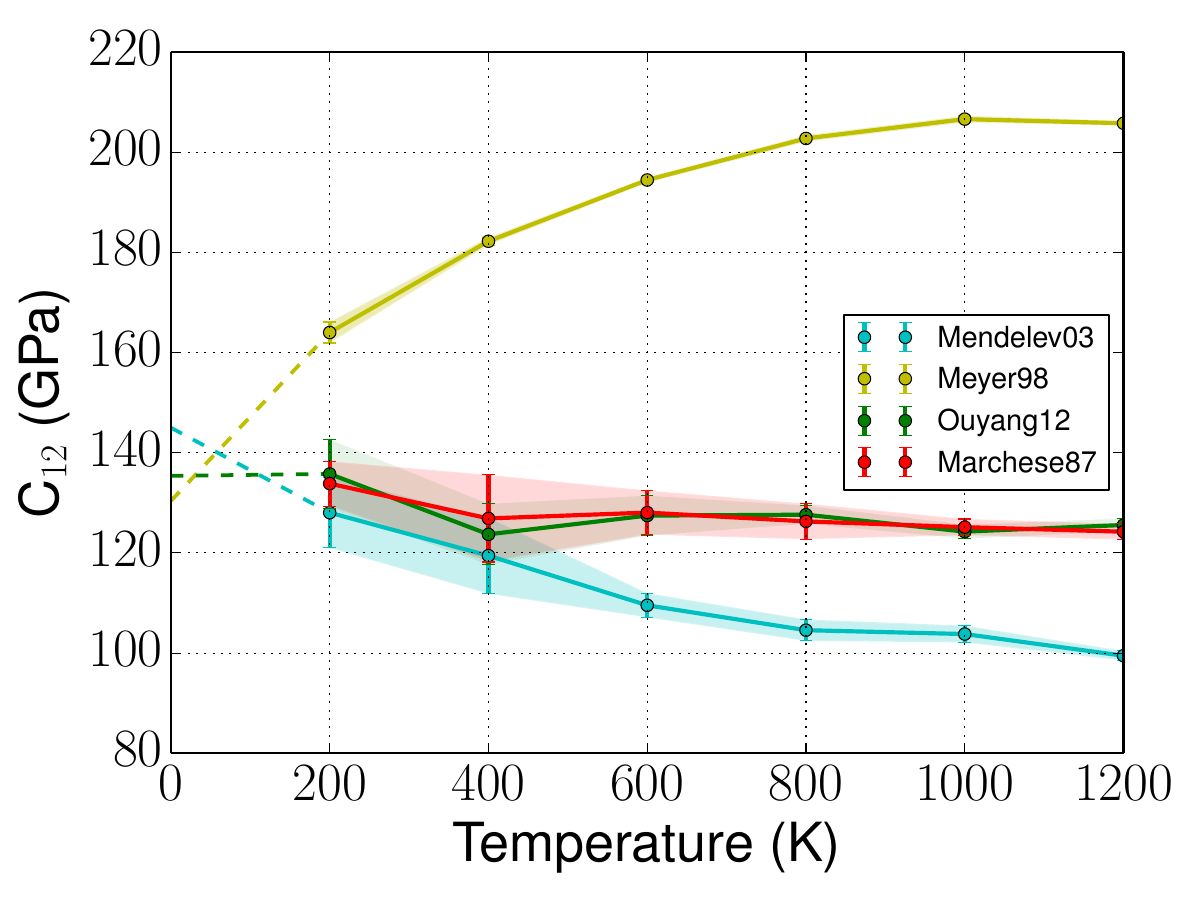} \\
   \includegraphics[trim=0mm 0mm 0mm 0mm, clip, width=0.46\textwidth]{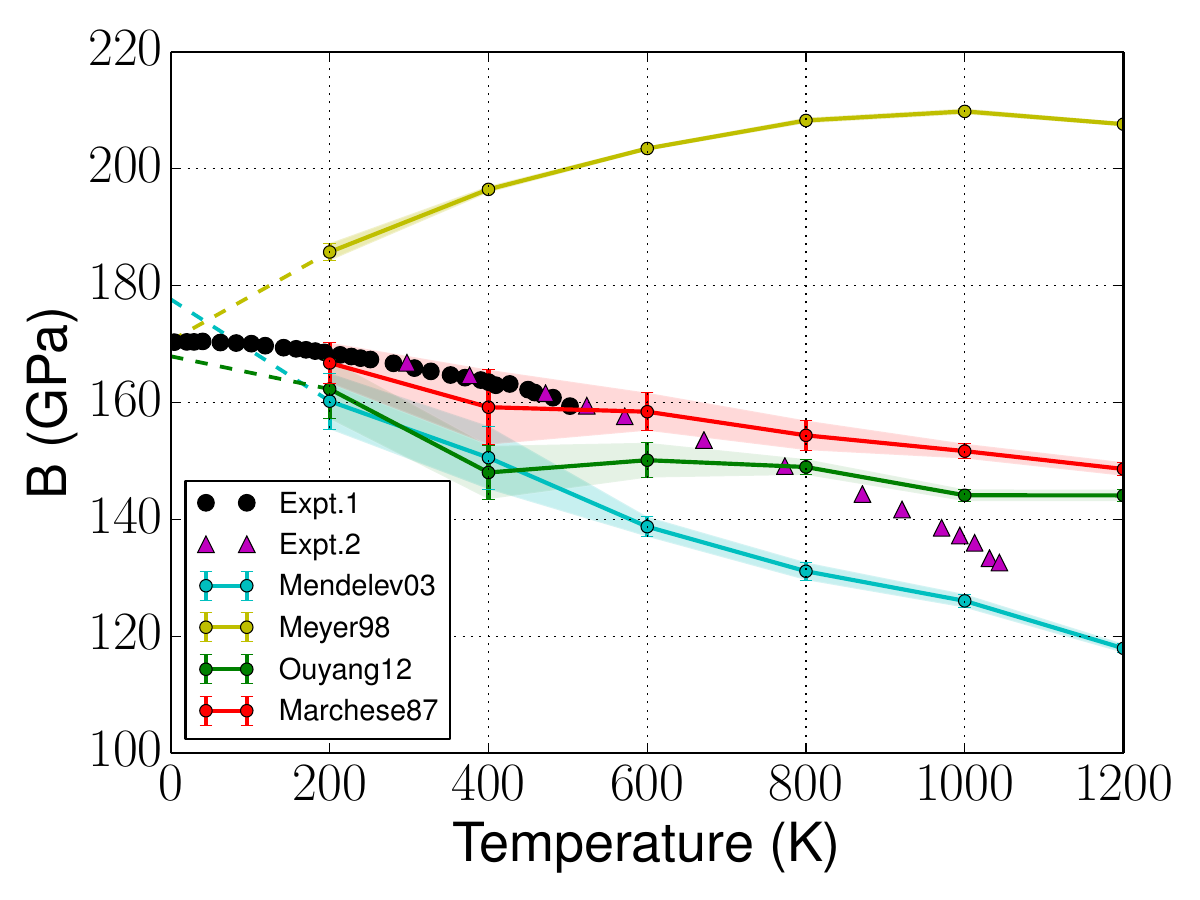} 
  \end{tabular}
 }
 \caption[...]{ Temperature dependence of the $C_{12}$ elastic constant (top panel) and bulk modulus $B$ (bottom panel) obtained from the different EAM potentials. The results are obtained as a linear 
 combination of those in Figs.~\ref{fig:C11_TS},~\ref{fig:Cp} according to the relations for cubic crystals in Sec.~\ref{sec:methods}. The ultrasonic experimental values from Ref.~\onlinecite{JJAdams} 
 (Expt.1 -- black circles) and Ref.~\onlinecite{Dever} (Expt.2 -- magenta triangles) are included as a reference.}
 \label{fig:C12B_T}
\end{figure}
%
In the second part of the work, we systematically survey the vibrational and thermoelastic properties of the potentials. 
We first calculate phonon dispersion at different temperatures. These are
reported and compared to experimental data at 300~K and 1158~K in Fig.~\ref{fig:PH_temperature}.
The Mendelev03 reproduces rather well the high-temperature phonon dispersions,
whereas the Marchese87 potential reproduces best the low-temperature data.
Meyer98 and Ouyang12 potentials yield too soft phonons at the $N$ point
and along the $\Gamma-N$ direction. Finally, we note that the softening of
the phonon modes from 200~K to 1200~K is best reproduced by Mendelev03.

The $C_{11}, C_{44}, C'$ elastic constants and their uncertainties are obtained from the sound velocities through the method described in Sec.~\ref{sec:checks}, making use of Eqs.~(\ref{eq:cost}), 
and are reported in Figs.~\ref{fig:C11_TS},~\ref{fig:Cp},~\ref{fig:C44}. For the sake of completeness, the $C_{12}$ elastic constant and the bulk modulus $B$ are also derived from standard relationships 
for cubic crystals and their values are reported in Figs.~\ref{fig:C12B_T}.

The extrapolated 0~K values are consistent with those originally calculated for the different potentials (see Tab.~\ref{tab:0K-volumes}) (without zero point correction) 
and with the experimental dataset that we report. This was expected, since all these potentials have been generated/fitted taking into account explicitly
the low-temperature experimental elastic constant values. The scattering of the zero temperature values 
associated to the various potentials is ascribable to the different experimental data used for the fitting and can be considered a measure of the experimental uncertainty. 
The results show also an overall deteriorating ability of the selected potentials in reproducing the experimental data upon increasing the temperature. In the high-temperature 
regime, far from the region in which they have been trained, large differences among the potentials themselves and large deviations with respect to experiments arise.
On average, the $C_{11}, C'$ and $B$ display the largest deviation from experiments, while better performance are obtained for the nearly linear $C_{44}$ temperature dependence. 
Interestingly, the $C_{44}$ elastic constant of the Mendelev03 displays a rapid unexpected softening with absolute values that are 18\% lower than the experiments already at 200~K.
A particular case is the Meyer98 that shows an unexpected marked stiffening of the $C_{12}$ and $B$ accompanied by a $C'$ that softens rapidly at low $T$ with a positive curvature.
This softening of the Meyer98 shear modulus $C'$ suggests a vanishing restoring force for atomic displacements along the [1$\overline{1}$0] direction associated to the long 
wavelength $T_2[110]$ normal mode, and a consequent possible mechanical/structural instability of the bcc structure towards an fcc through a Bain path transformation~\cite{Neugebauer-phonon,Petry}.
Noticeably, as discussed in Ref.~\onlinecite{engin}, Meyer98 is one of the few EAM potentials that allows for a thermodynamic stabilization of the fcc phase with respect to the bcc one upon increasing 
temperature. In particular, the Meyer98 free energy for the fcc phase is lower than the bcc phase already at 600~K, and the softening of $C'$ can be considered as a precursor mechanism for the 
bcc/fcc transformation.

As reported by various studies in the literature~\cite{Hasegawa,Rusanu,Dever,Dragoni}, we expect magnetic disorder effects become increasingly important in the description of 
the marked non-linear softening of some elastic constants approximately above two-thirds of the Curie point. 
The poor agreement of the calculated curves with the experimental high-temperature behavior suggests therefore the difficulty for the EAM models considered to reproduce such effects. 
This mismatch at high temperature is expected since the standard EAM formalism does not contains any term to deal explicitly with magnetic degrees of freedom and no (or partial)  
high-temperature ferromagnetic and paramagnetic properties are included in the training protocol of the potentials. In order to address these deficiencies one could extend the training 
set of a standard EAM model to include paramagnetic or high-temperature ferromagnetic data, aiming at effectively reproducing the properties of interest. 
A second approach, that would allow to gain more physical insight into the study of those quantities dominated by magnetic disorder (for example the heat 
capacity~\cite{Neuge_C_P,Ma}), is based on the generalization of the EAM formalism to deal with magnetic degrees of freedom. 
This can be achieved 
at a semi-classical level by means of generalized spin-lattice dynamics~\cite{Ma_Woo_Dudarev,Pan2017}.

\section{Elastic constants and the phonons long wavelength limit}
\begin{figure}[!ht]
\centering{
  \begin{tabular}{c}
   \includegraphics[trim=0mm 0mm 0mm 0mm, clip, width=0.46\textwidth]{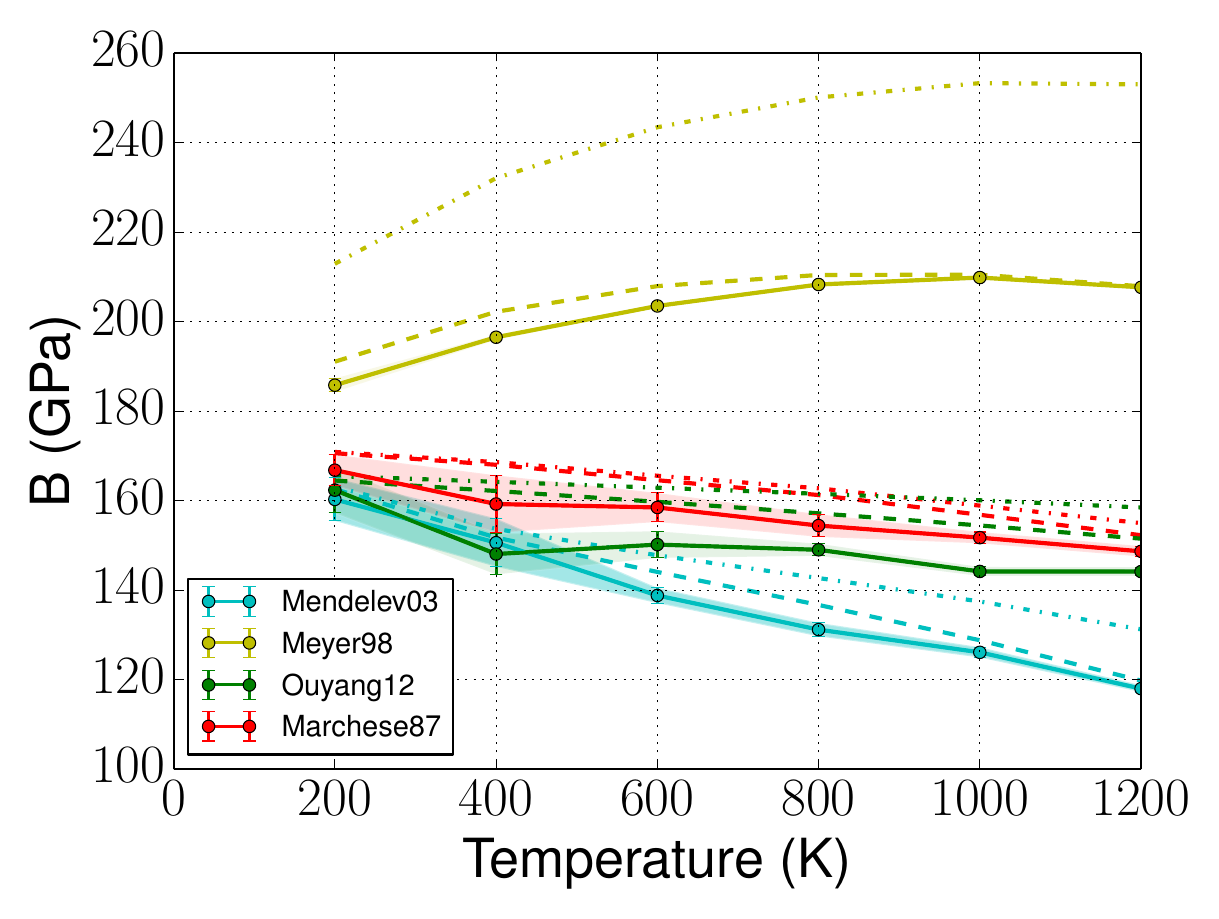} 
  \end{tabular}
 }
 \caption[...]{ Bulk modulus as a function of temperature for the four EAM potentials considered in this work. The solid lines correspond to the bulk modulus computed from the slope of the phonon dispersions. 
 These curves are compared with isothermal (dashed lines) and adiabatic (dash-dotted lines) bulk moduli obtained from a finite deformation approach combined to MD NVT runs. As mentioned in the main text, the approximation of considering the elastic constants obtained from the phonon slopes at finite temperature as either 
 isothermal and adiabatic constants does not hold for the Meyer98 case, where the adiabatic curve is far from the isothermal one. Such deviation is due to the specific anomalous thermal expansion and thermal expansion coefficient displayed by the potential. }
 \label{fig:B_comparison}
\end{figure}

As shown by Wallace~\cite{Wallace} in perturbation theory to second order in the phonon-phonon anharmonicity, the long-wavelength phonon waves at finite temperature propagate in a crystal neither as 
isothermal nor as adiabatic thermoelastic waves. 
The elastic constants that we extract from Eq.~\ref{eq:cost} and present in Sec.~\ref{sec:results} 
come from the phonon spectrum at finite temperature and, as such, have to be considered only approximations to either isothermal and adiabatic elastic constants defined form a thermodynamic point of view. 
In Fig.~\ref{fig:B_comparison}, we numerically demonstrate that this is a reasonable approximation for the bulk modulus (the only exception is the adiabatic bulk modulus of the Meyer98 potential), and 
we assume that it holds also for the other elastic constants~\footnote{Note however that for the $C_{44}$ and $C'$ shear elastic constants there is no difference between isothermal and adiabatic propagation}. 
The isothermal curves reported in Fig.~\ref{fig:B_comparison} are obtained from a finite deformation approach combined to MD NVT runs. At each temperature, the equilibrium volume is changed 
hydrostatically in a range  from [-3,+3]\% and the Cauchy stress is calculated as the  time average of the virial stress estimator including the kinetic term~\cite{zimmerman}. The linear coefficient 
of proportionality of  pressure and volume is taken as isothermal bulk modulus. The isothermal condition stems from  the NVT ensemble used in the MD calculations. 
The details of the MD calculations are the same discussed in Sec.~\ref{sec:methods} for the calculation of the equilibrium volumes.
The adiabatic correction to the isothermal bulk modulus $B^{(T)}$ is instead obtained from the following relation:
\begin{align}
B^{(S)} - B^{(T)} &= \frac{T V \alpha_V^2 {B^{(T)}}^2}{C_P -TV\alpha_V^2B^{(T)}}  .
\label{eq:adia_correction}        
\end{align}

\section{Conclusions}
\label{sec:conclusions}
We calculated the thermal expansion and thermal expansion coefficient, the heat capacity at constant pressure, the phonon spectra and the elastic constants as a function of temperature of 
four embedded-atom method potentials. We focused on the bcc phase of iron ranging from 0 to 1200~K (the experimental temperature range of of the $\alpha$ and $\beta$ phases). 
Our calculations were based on classical molecular dynamics, thus accounting for the phonon-phonon anharmonic contributions that are assumed to dominate the phonon frequencies renormalization at high temperature, 
while neglecting quantum statistical effects at low temperature. The elastic constants were obtained from the slope of the phonons in $\Gamma$ at finite temperature. As such they are interpreted as a 
mechanical approximation to the elastic constants defined thermodynamically. 
The performance of the potentials considered herein are compared to experiments and, when available, to ab-initio data from quasi-harmonic theory.
The results show strengths and weaknesses of the models as the temperature is raised and suggest that none of the models yields convincing thermodynamic properties.
The discrepancies with respect to experiments in the low/intermediate regime are justified with the limited ability of the specific models to accurately describe the phonon softening. 
In particular we found anomalies in the softening of the $C_{44}$ of the Mendelev03 potential and in the $C',C_{12}$ and $B$ of the Meyer98 potential. 
These should be taken into account when studying finite temperature mechanical properties of both elemental iron and iron alloys with these models. 
The lack of magnetic degrees of freedom in the EAM formalism is also expected to contribute in explaining the increasingly large deviations from experiments in the 
high-temperature regime. 

\section{Acknowledgements}
The authors thank Dr. Matteo Cococcioni for the helpful discussions. They also gratefully acknowledge the financial support from the Swiss National Science Foundation (SNSF – Project No. 200021–143636) and 
partial support from the Seventh Framework Programme MINTWELD Collaborative Project. This work was supported by a grant from the Swiss National Supercomputing Centre (CSCS) under project ID \textit{s337}.

\bibliography{EAM5}

\end{document}